\documentclass[preprint,journal]{vgtc}       

\ifpdf
  \pdfoutput=1\relax                   
  \usepackage{graphicx}                
  \DeclareGraphicsExtensions{.pdf,.png,.jpg,.jpeg} 
\else
  \usepackage{graphicx}                
  \DeclareGraphicsExtensions{.eps}     
\fi%

\graphicspath{{figures/}{pictures/}{images/}{./}} 

\usepackage{microtype}                 
\PassOptionsToPackage{warn}{textcomp}  
\usepackage{textcomp}                  
\usepackage{mathptmx}                  
\usepackage{times}                     
\usepackage{tabu}                      
\usepackage{booktabs}                  
\usepackage[utf8]{inputenc}
\usepackage[english]{babel}
\usepackage{amsmath}
\usepackage[normalem]{ulem}
\usepackage{float}
\usepackage{color}  
\usepackage{tikz}  
\usepackage{pgfplots}
\usepackage{adjustbox}
\usepackage{cleveref}
\usepackage{macros}
\usepackage{relsize}
\usepackage{pgfplots}
\newcommand{\rowgroup}[1]{\hspace{-1em}#1}

\pgfplotsset{every axis/.append style={
                    label style={font=\tiny},
                    tick label style={font=\tiny} ,
                    tick label style = {font=\fontsize{6 pt}{6 pt}},
                    every axis label = {font=\fontsize{6 pt}{6 pt}},
                    legend style = {font=\roboto},
                    label style = {font=\fontsize{7 pt}{7 pt}}
                    }}

\usepackage[english]{babel}

\usepackage[square,sort,numbers]{natbib}

\usepackage{multirow}
\pgfplotsset{compat=newest}

\onlineid{1198}

\vgtccategory{Research}
\vgtcpapertype{theory/model}

\title{Visualization Equilibrium}

\author{Paula Kayongo, Glenn Sun, Jason Hartline, and Jessica Hullman}
\authorfooter{
\item
Paula Kayongo is with  Northwestern University. E-mail: paulakayongo2023@u.northwestern.edu
\item
 Glenn Sun is with University of California, Los Angeles. E-mail: glennsun@g.ucla.edu.
 \item
 Jason Hartline is with Northwestern University . E-mail:hartline@eecs.northwestern.edu.
\item
 Jessica Hullman is with Northwestern University. E-mail: jhullman@northwestern.edu.
}

\shortauthortitle{Biv \MakeLowercase{\textit{et al.}}: Global Illumination for Fun and Profit}

\abstract{%
In many real-world strategic settings, people use information displays
to make decisions. In these settings, an information provider chooses
which information to provide to strategic agents and how to present
it, and agents formulate a best response based on the information and
their anticipation of how others will behave.  We contribute the
results of a controlled online experiment to examine how the provision
and presentation of information impacts people's decisions in a
congestion game. Our experiment compares how different visualization
approaches for displaying this information, including bar charts and
hypothetical outcome plots, and different information conditions, including
where the visualized information is private versus public (i.e., available
to all agents), affect decision making and welfare. We characterize
the effects of visualization anticipation, referring to changes to
behavior when an agent goes from alone having access to a
visualization to knowing that others also have access to the
visualization to guide their decisions. We also empirically identify
the visualization equilibrium, i.e., the visualization for which the
visualized outcome of agents' decisions matches the realized decisions
of the agents who view it. We reflect on the implications of
visualization equilibria and visualization anticipation for designing
information displays for real-world strategic settings.
}

\keywords{Visualization equilibrium, Uncertainty visualization, Strategic communication, Nash equilibrium.}

\CCScatlist{ 
 \CCScat{K.6.1}{Management of Computing and Information Systems}%
{Project and People Management}{Life Cycle};
 \CCScat{K.7.m}{The Computing Profession}{Miscellaneous}{Ethics}
}

\begin{document}


\maketitle

\section{Introduction}
In real-world strategic settings, people use information displays to
make decisions. For example, data visualizations are
used to communicate payoff-relevant information to agents in settings
like Internet advertising auctions through dashboards like Google's
AdWords, which advertisers use to determine how much they will bid to
obtain an advertising spot on a page or a keyword. Visualizations are a natural way to present payoff relevant information in strategic settings because of their abilities to offload cognition to perception~\cite{larkin1987diagram} and making information more salient~\cite{jarvenpaa1990graphic}. 

The design of visualization displays for strategic settings introduces
new considerations for developers. Consider a strategic setting where multiple agents use the same display to make decisions. The developer of such a display might realize the predictions it reports will affect individual decisions, leading to distribution shifts due to these responses thereby decreasing the accuracy of the predictions.  A natural approach for this developer would be to adopt the canonical Nash equilibrium concept from game theory and make the prediction that corresponds to a Nash equilibrium.\footnote{A Nash equilibrium for a strategic scenario prescribes a strategy for each agent with the property that each agent is happy to follow her prescribed strategy if all other agents do so as well.  Specifically, her payoff for this
strategy is at least as good as any other strategy.  Simulating the
strategies, a Nash equilibrium gives a prediction of the outcome to
expect.} If all agents Nash best respond to the Nash prediction, by selecting a choice that results in the most favorable individual outcome given the actions of others, then such a prediction would be correct.  However, research in visualizaton indicates that many choices involved in specifying a chart---such as marks, encodings, annotations, uncertainty depictions, and data sampling---impact decisions made from that visualization.
Does visualizing a Nash equilibrium outcome result in this best
response behavior? We conduct a crowdsourced experiment to show that the answer is ``no''. Our mixed design experiment explores the relationship between visualized behavior and resulting behavior for a game where payoffs depend on one's own choice and the choices of others playing the game at the same time.

We develop the concept of a visualization equilibrium to address the
developer’s problem and use our experiment results to evaluate
equilibria across different combinations of information and
visualization strategies. One can think of a visualization equilibrium
as adjusting the visualized estimates to account for behavioral reactions to the visualization so that these visualized estimates correspond to the
aggregate behavior under the visualization. Visualization equilibrium is important in applications like internet advertising auctions where any visualization other than the equilibrium
is incorrect and could result in unpredictable dynamics or users
losing trust in the application. Using a grid search over visualized binomial distributions for those that predict their own realized outcomes, we identify the visualization equilibrium for several visualization types. Specifically, we study static versus animated hypothetical outcome bar charts, which vary in how salient they make uncertainty about behavior. We find that the visualization equilibria are
similar between them, but both are different from the Nash equilibrium.

Two factors contribute to an agent's response to a
visualization in a strategic setting: their understanding of the
visualization and their understanding of other agents'
understandings.  Our experiment separates these two factors of strategic
visualization understanding by considering public and private
visualization conditions.  The inclusion of this private visualization condition allows us to study the role of \textit{visualization anticipation}, the difference in the distribution of actions that results in a setting where the visualization is public rather than private. We find that when the visualized information is public, players were only two thirds as likely to make the decision that would result in a higher payoff according to the visualization. 

Additionally, anticipation may vary with different kinds of visualizations. We speculated that visualizations that make data uncertainty more salient may make it more challenging for agents to anticipate others' reactions. However, we find that while overall, a more salient presentation of uncertainty (HOPs~\cite{hullman2015}) makes participants less likely to select the higher payoff location, the interaction between visualization approach and whether the visualization is public and private is slight. While a less ambiguous visualization (static bar charts) leads to slightly more distinct behavior comparing between public and private visualization access relative to HOPs, this effect is not reliable.  



We use qualitative analysis of reported strategies to provide additional context on how participants made strategic decisions when a visualization of payoff-relevant information was public.  We report on how frequently participants describe explicitly considering other players' behavior, responding randomly in the absence of a clear best response, best responding to the visualization alone, and responding to the payoff function independent of the visualization. We discuss the implications of our results on designing displays for strategic settings and directions for future work.

\section{Background and Motivation}

\subsection{Using visualization to coordinate behavior}
Research in information visualization studies how visualizations can be used to coordinate the behavior of multiple people. Social visualization systems like sense.us~\citep{heer2009senseus}, BabyNameVoyager~\citep{wattenberg2005baby}, and CommentSpace~\citep{Wesley2011commentspace} incorporate social signals of others' behavior to support collaborative sensemaking. ~\citet{willet2007scented} found that
social features like scented widgets, which visualize traces of past behavior with a visualization system to help new users navigate to potentially interesting views, help users make twice as many unique discoveries with unfamiliar datasets.  However, they caution that traces of social information may hamper a group of analysts' coverage of the data set, precluding some discoveries. Another study found that access to social information about prior users' graphical judgments can bias subsequent users' perceptions of data~\cite{hullman2011impact}. Our work differs from the prior work on social visualization in that while we display prior behavior using visualizations, we study settings in which the visualization is intended to be an accurate depiction of the behavior that results from it, rather than just a depiction of past activity. 
We pursue a formal approach to defining coordination via a shared visualization as visualization equilibrium, the point where a visualization perfectly predicts the resulting behavior of agents who view it. We also study the role of anticipating others' reactions to a shared visualization on behavior in strategic settings with a well-defined payoff structure.

\subsection{Decision-making under uncertainty with visualization}
Depiction of uncertainty may play a role in settings where people make strategic decisions from shared displays. In such settings presenting only point estimates can imply ``incredible certitude''~\citep{manski2018} where people assume the visualized estimates are more precise than is warranted. Communicating uncertainty alongside point estimates, including by visualizing it, has been found to improve decision making in various settings where displays present weather and transit (e.g.,~\citep{fernandes2018,SusanJoslyn}) information. However, uncertainty displays can make estimates more ambiguous and therefore harder to anticipate reactions to~\cite{hullman2019}, or bias interpretations in ways that impact strategic decisions being made~\cite{westwood2020projecting}. Most studies of uncertainty visualization explore individual decision making in non-strategic settings~\citep{fernandes2018,KaleAlex2020VRSf,Bancilhon2020LetsGH}. In contrast, our work examines whether different representations of distributional information (i.e frequency representations like animated hypothetical outcome plots (HOPs)~\citep{hullman2015} versus static bar charts) affect judgments and incentivized decisions in non-strategic versus strategic settings. 

\subsection{Persuasion}

\subsubsection{Visualization Persuasion}
Researchers have characterized and empirically studied the persuasive affordances of visualizations. \citet{hullman2011visualization} taxonomized forms of visualization framing effects achieved by rhetorical decisions authors make to shift viewers' beliefs about data, such as choices about which information to make accessible to viewers or how to convey provenance information to build viewers' trust. In a controlled experiment, \citet{pandey} found that visual information can be more persuasive than textual information on viewers' beliefs, however, the extent to which the visual information was persuasive was modulated by the degree to which individuals had an opinion on the subject.
\citet{correllBlackHat} described black hat visualization as intentionally misleading use of visual representations of data. More formally, \citet{kindlmann2014algebraic} use algebraic considerations related to visualization such as the relationship between the mathematical structure of the data, its visual representation, and its perception by humans to define a model of visualization design. Their work identifies formally defined principles of effective visualization like representation invariance, unambiguous depiction of data, and visual data correspondence. \citet{correll2018looks} use the algebraic model to study how adversarial or even simply careless setting of design parameters of visualizations such as histogram bin widths can obscure important flaws in data. Our work takes a slightly broader view of persuasion through visualization by considering how different combinations of payoff-relevant information and visualization approach can shift aggregate-level outcomes like equilibria in strategic settings. Our ultimate goal is to understand which visualization choices support more stable, predictable, and rational behavior in strategic environments.  

\subsubsection{Information Design (in Economics)}
In economics, information design~\citep{dirkbergemann2019infosedign}
considers situations where an information provider (principal) can
commit to providing payoff relevant information (signal) about possible states of the world to agents who are making decisions.
The goal of the principal is to provide information that results in
the agents making decisions that are better for the principal.
Information design can be applied in the classical economic model of
Bayesian games~\citep{harsanyi1968games} where agents share a common
prior belief over the states. In this application, the principal can
send the agents signals that are correlated with the state. The
agents do Bayesian updating and act according to their posterior
beliefs on the state, other agents' beliefs, and other agents' strategies.
This literature is primarily concerned with identifying signaling
schemes that are optimal from the principal's perspective. Most
related to the congestion game considered in this paper,
\citet{DasSanmay2017Rcti} provide theoretical analysis of a model of
road networks that demonstrates the benefits of information design.
We adopt a similar perspective to information design with three major
differences.  First, we do not assume that agents are able to
perfectly optimize their action from their received signal.  Second,
we allow that different types of visualizations of the same signal may
result in different behaviors (e.g., point estimates shown as bars,
hypothetical outcomes drawn from a distribution).  Third, to focus on
the previous two behavioral differences from the theoretical models,
we consider a setting, which would be trivial for theoretical
information design, where the agents uncertainty is only on the
behavior of other agents (and not on a payoff relevant state).

\subsection{Performative Prediction}
A challenge can arise in deploying model predictions of behavior when
the behavior is responsive to the predictions.  The behavior induced
by the prediction may be different from the behavior in the training
data absent a prediction.  The task of encorporating this behavioral
response in the prediction has been termed performative
prediction
and is studied in the machine learning literature~\cite{perdomo2020performative,brown2020performative,miller2021outside}. Performative
prediction is relevant in application areas like voting where polls
may affect voting decisions and map navigation where traffic
predictions may affect route planning and thus the realized
traffic~\cite{buscema2009impact}.  In addition to providing
theoretical insight into when a fixed point within a repeated model
retraining process can be found (i.e., a model which would prevent
having to retrain)~\cite{perdomo2020performative}, recent research
describes optimization procedures that can be used to minimize this
shift from behavioral
responses~\cite{izzo2021learn,miller2021outside}. Our work takes a
different approach by considering the role of the interface, such as a
visualization, in rendering predictions of behavior
inaccurate. Similar to the ideas of a model that minimizes the
difference between a predicted and induced distribution, we present
the notion of a visualization equilibrium, the point at which the
visualization achieves minimal error with respect to the induced
distribution.


\section{Equilibria and Welfare}
To understand visualization use in strategic decision making, we study a congestion game commonly studied in game theory.  Below we define the
game, demonstrate a Nash equilibrium, and the optimal (i.e.,
highest obtainable) social welfare.

\subsection{Formal Game Set Up}

Congestion games are a class of non-cooperative games where the payoff
of an agent's action is determined by their selection as well as the
number of other agents who select the same congestible element
\citep{Rosenthal1973ACO}. Congestion games provide a simple model
under which to understand selfish behavior in strategic environments, and model several real-world problems in which agent utility depends on the total number of agents making a similar decision, such as traffic congestion routing on communication and
road networks \citep{RoughgardenTardos2002}, spectrum sharing\citep{mliu2009}, and firms selecting between alternative production
processes \citep{Rosenthal1973ACO}.\footnote{The narrative presented to participants in our experimental study most closely resembles that of
The El Farol Bar problem \citep{elfarol} which models how a fixed
population selects to go to a bar to minimize crowding.}

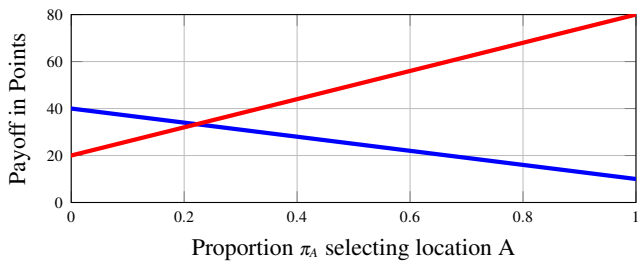
\begin{figure}[t]
 \centering
  \begin{tikzpicture} 
    \begin{axis}[width=.85\columnwidth ,height=2.5cm,
    enlargelimits=false,scale only axis=true, xmin=0, xmax=1,ymin=0, ymax=80,samples=50,grid=major,xlabel={Proportion $\prA$ selecting location A },ylabel={Payoff in Points}]
    \addplot[blue, ultra thick](x,{40-(30*x)});
    \addplot[red, ultra thick](x,{80-2*(30*(1-x))});
     \end{axis}
  \end{tikzpicture}
    \caption{Payoff at location A and B as a function of the
      proportion $\prA$ of agents selecting location A. The payoff at
      location A and B is the same at $\prA = 0.22$. }
    \label{fig:payoff}
\end{figure}

We focus on a simple two-choice
congestion game with an affine payoff function. The game is non-atomic, meaning the number of agents is high enough to be approximated by a continuum. The agents choose between two locations A and
B.  The decision of this continuum of agents can be summarized by
$\prA \in [0,1]$, the fraction of them that choose location A,
leaving a $\prB = 1-\prA$ fraction choosing location B.  The affine
payoff function for the agents choosing A and B, respectively is
(\Cref{fig:payoff}):

\begin{align*}
  \payoffA(\prA) &= 40 - 30\, \prA
 &
  \payoffB(\prA) &= 80 - 60\, (1-\prA)\\
  &&& = 20 + 60\, \prA 
  \tag{1}\label{eq1}
\end{align*}

We choose this model because it exhibits
typical behavior for a congestion game (as we expand on below) and it
is simple to communicate to untrained participants in our experimental
study.

\subsection{Nash Equilibria}
\label{sec:nash}

The field of game theory suggests Nash equilibrium as a behavioral
prediction for rational agents in strategic situations. Nash
equilibrium is a profile of strategies (one for each agent) such that
every agent is happy to follow their strategy given that all other
agents are following their strategies.  

Our congestion game admits a unique Nash equilibrium when a $\prAnash =
2/9 \approx 0.22$ fraction of the agents choose location A.  Specifically,
with this fraction choosing location A, the payoffs of location A and
B are equal, and all agents are indifferent with their choice.  There are
two main ways to achieve such an equilibrium.  A specific $2/9$ fraction of the
agents could deterministically choose location A, while the other $7/9$ fraction
deterministically choose location B.  Alternatively, all agents could
randomly pick A with probability $2/9$, and B with the remaining
$7/9$ probability.  It is a fundamental property of non-atomic
congestion games that the equilibrium payoff of every action that is
played at equilibrium is the same ~\cite{Roughgarden2004}. 

\subsection{Social Welfare}
\label{sec:social}

 
 
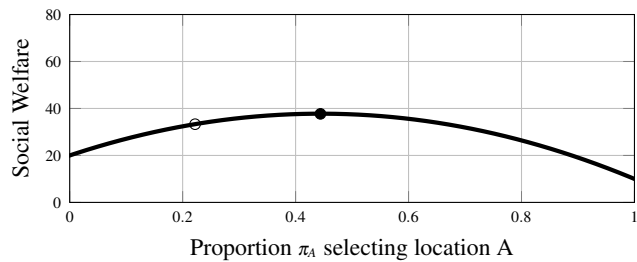
\begin{figure}[t]
    \centering
   \begin{tikzpicture}
    \begin{axis}[ width=.85\columnwidth ,height=2.5cm,
    enlargelimits=false,scale only axis=true,xmin=0, xmax=1,ymin=0, ymax=80,samples=500,grid=major,xlabel={Proportion $\prA$ selecting location A },ylabel={Social Welfare}]
      \addplot[black, ultra thick](x,20+80*x-90*x^2);
      \addplot[mark=*] coordinates {(0.444,37.7)}; 
      \addplot[mark=o] coordinates {(0.222,33.3)}; 
    \end{axis}
    \end{tikzpicture}
    \caption{Social welfare as a function of the proportion of people
      selecting location A.  Social welfare is maximized with a
      $\prAopt = 4/9 \approx 0.44$ fraction of the agents choosing
      location A (marked by $\bullet$).  The Nash equilibrium is at
      $\prAnash = 2/9 \approx 0.22$ (marked by $\circ$). }
    \label{fig:socialwelfare}
     \vspace{-5mm}
\end{figure}

 An analysis of a congestion game is often concerned with social welfare, i.e., the average payoff obtained by the agents. The
social welfare of our congestion game can be easily expressed as a
function of the fraction $\prA$ of agents that choose location A
(\Cref{fig:socialwelfare}).  The social welfare is maximized at
$\prAopt = 4/9 \approx 0.44$.

\begin{align*}
  \SW(\prA) &= \prA\ \payoffA(\prA) + (1-\prA)\,\payoffB(\prA)\\
  &= 20 + 80\,\prA - 90\,\prA^2
   \tag{2}\label{eq2}
\end{align*}

 Social welfare sums the utilities that all players obtain in the game. When social welfare is maximized society has achieved the most optimal outcome as a population. Naturally, agents presented with visualizations of congestion may not
act according to the Nash equilibrium, because, for example, they are
imperfect information processors, or have imperfect strategic
thinking. Not playing in Nash equilibrium, however, is not necessarily bad
from the perspective of overall welfare. Below, we show through our experiment how the welfare at visualization equilibrium of our congestion game is more socially optimal than the welfare at  Nash equilibrium.





\subsection{Visualization Equilibrium}

The visualization equilibrium is the visualization for which the
observed behavior mimics or approximately mimics the visualized
behavior. Speaking mathematically, a visualization equilibrium is a
fixed point of the composition of the agents' behavior and the
function that produces the visualization.

We consider visualizations based on small samples of data,
specifically $N=30$ decisions of prior agent behavior.  Given a true
rate $\prA$ that a prior agent selects location A, the sampled data is
a binomial distribution with non-trivial variance.\footnote{Note that
  while $N=30$ agents is well approximated by a continuum in game
  theoretic analysis, $N=30$ samples is a small sample, i.e., not well
  approximated by a continuum for statistical analysis.}  

In our congestion game, a visualization is provided as a summary of
past play that can be used as a prediction of the aggregate outcome of
the game. To make a binary decision between two locations, the first moment (mean) is sufficient. However, to estimate how probable it is that one location will yield a higher payoff requires considering the sampling error affecting the statistic. We consider different visualizations of uncertainty in our study under the expectation that making uncertainty around the true rate more salient might affect coordination by making it harder to anticipate other participants' reactions to the visualization.

Recall that outcomes in our congestion game are
summarized by $\prA$, the fraction of agents choosing location A, with
the fraction $\prB$ selecting B equal to $1-\prA$. Given a visualization $\VIS$ comprised of a visualization specification with data defined as $\prApred$ and definitions of marks, encodings, scales, references, etc., the strategies of the agents result in some outcome. The visualization
equilibrium of a visualization scheme is the prediction for which the
outcome resulting from the visualization is the same as the
prediction.  Formally:
\begin{itemize}
\vspace{-2mm}
\item A visualization $\VIS$ maps a predicted outcome $\prApred$ to a visualization $\VIS(\prApred)$.
\vspace{-3mm}
\item Aggregate agent strategies $\STRAT$ map a visualization $\VIS$ to an outcome $\STRAT(\VIS)$.
\vspace{-3mm}
\item A visualization equilibrium $\prAviseq$ is defined as satisfying $\prAviseq = \STRAT(\VIS(\prAviseq))$, i.e., it is a fixed point for $\STRAT(\VIS(\cdot))$.
\vspace{-2mm}
\end{itemize}

We assume aggregate agent strategies $\STRAT$ to be given by the environment.
Varying the non-data aspects of $\VIS$'s specification, such as by choosing a different set of marks and encodings or adding encodings, as we do in our experiment, may result in different visualization equilibria and thus result in different social welfare. When this is the case, the information provider or principal may prefer visualizations with higher welfare at the visualization equilibrium. 

\section{Experimental Method}
\begin{figure}[t]
 \centering
  \includegraphics[width=\linewidth]{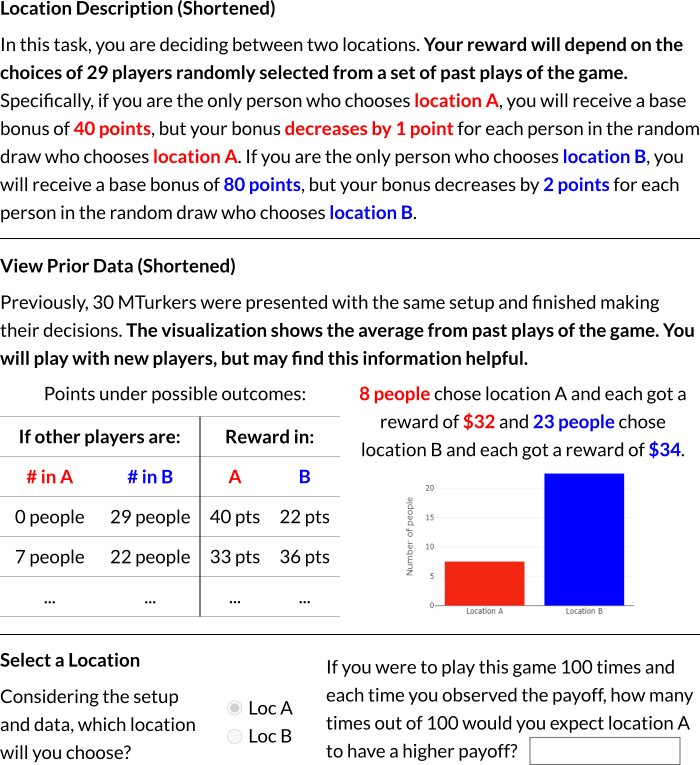}
   \caption{Depiction of the study interface. The example shows the task in a private visualization with static bar charts, where the participant has already selected location A. The text and table above are condensed. Participants also received additional text information about the setup and data.}
   \label{fig:interface}
 \end{figure}

We designed a mixed design repeated measures experiment to answer two research questions:

\noindent\textbf{R1:} What \textit{information signalling strategy} results in a visualization equilibrium, where the visualized behavior accurately reflects the realized behavior of decision-makers who view that visualization?

\noindent\textbf{R2:} How do decisions and judgments under uncertainty change when visualizations conveying payoff-relevant information are accessible only to a single individual (\textit{private visualization}) versus when the visualizations are available to all agents making decisions (\textit{public visualization})?

R1 helps us understand, for a particular congestion game, the contribution of choices related to information signal and visualization approach on how agents ``play.'' R2 helps us examine the role of visualization anticipation, the change in decisions that agents make based on their anticipation of other agents' reactions to the same display.

\subsection{Experimental Tasks and Incentives}
In our experiment, players make a series of independent binary decisions (constituting trials) about which of two locations, A or B, to visit. We frame the decision as a selection between one of two locations (e.g., beaches), where the payoff at each location is a function of the number of people selecting the location. Participants are provided with information about the payoff structure (equation \ref{eq1}) and a conversion table that helps them map different possible outcomes to different payoffs they would receive (Figure \ref{fig:interface}). Each trial is characterized by an \textit{information signalling strategy}  defined by a \textit{visualization type} (HOPs or static bar chart), the presence or absence of prior information relevant to the game (\textit{information presence}), whether the information is provided to only one participant or to all participants (\textit{information access}), and what information is visualized (\textit{information signal}), which in our experiment is the location choice of 30 prior players.  For each trial, our experiment also asks participants to report their estimate of the probability that the location they selected results in a higher payoff (Figure \ref{fig:interface}).  

\subsection{Experimental Design}

\begin{figure}[t]
    \centering
    \includegraphics[width=\columnwidth]{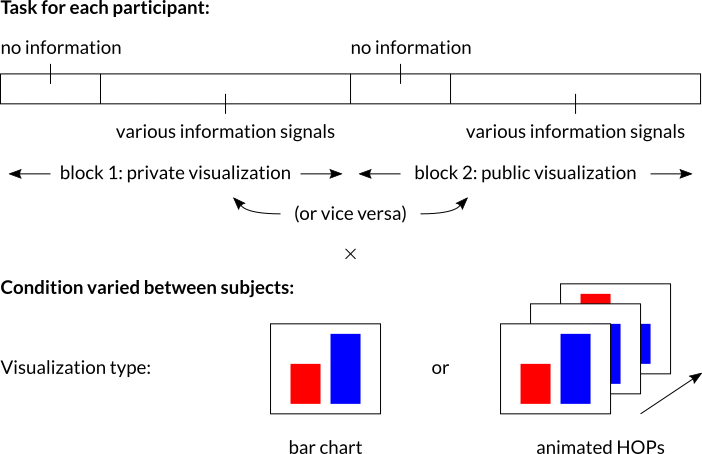}
    \caption{Diagram of procedure (top) and visualization manipulation. Information presence, access, and signal were varied within subjects. Visualization type was varied between subjects.}
    \label{fig:expdesign}
\end{figure}

Our experiment varied the following factors: 

\noindent \textbf{Information Presence}: 
To understand how participants behave in the absence of prior information, we vary whether or not participants are provided with information on prior plays of the game. In trials where information was not provided through visualization, participants had access to only the payoff structure of the game.  

\noindent \textbf{Information Access}: 
In conditions where prior information was provided, we varied whether the prior information was framed as available to all others making the decision (\textit{public visualization}) or whether it was only available to the participant (\textit{private visualization}). In the public visualization condition, participants were informed that their reward will depend on the decisions of other participants who similarly have access to the visualization. In the private visualization condition, participants were informed that their reward would depend on the actions of a subset of randomly drawn participants from past play who did not see the visualizations when making their decisions. They are also informed that the visualizations are representative samples from the subsets that will be used to compute their reward for each trial. We explicitly varied information access for the empirical purpose of measuring the effect of anticipation on decision making.


\noindent \textbf{Information Signal}: In trials where information is present, we vary the proportion shown in the visualization. In particular, we visualized proportions of people choosing location A between 0.1 and 0.5, in increments of 0.05 across nine trials. We selected the range of probabilities to be roughly centered around $\prAnash \approx 0.22$ because for our payoff function (equation \ref{eq1}), location A is preferable to location B when the proportion $\prA$ going to A satisfies $\prA < \prAnash$, and the relationship is inverted when $\prA > \prAnash$.

\noindent \textbf{Visualization Type}: In trials where prior information is presented, participants view a visualization showing the number of players who selected location A and location B in past plays of the game (\Cref{fig:interface}). In one visualization condition, participants view static bar charts of the proportion of prior participants choosing location A versus B. In the other condition, participants view HOP bar charts where each frame depicts a random draw from a binomial distribution with $p$ set to the proportion assigned for that trial. By encoding uncertainty via the same bar encoding but through temporal frequency, which is known to be automatically processed by people~\cite{hasher1984automatic}, we expect HOPs to make the uncertainty more difficult to ignore compared to static bars~\cite{hullman2015}. Through temporal frequency, HOPs also provide information about the probability that a location selection is the higher payoff location, which is directly relevant to answering the probability judgment question in the private trials. In both cases, participants were informed that the visualization depicted past play; for HOPs, participants were informed that the animation displayed a set of random draws from a past play of the game.

We use a counterbalanced blocked design with two blocks, one of ten public visualization trials, the other of ten private visualization trials. The first trial in each block was the no-information trial, followed by the nine information signals in random order. The two trials where no prior information was presented varied slightly in their description between blocks to match the decision circumstances under private versus public information. Specifically, the no-information trial in the private information block informed participants that they would be making decisions against a random subset of players drawn from past play. The no-information trial in the public information block informed participants that they would be making decisions against current players similarly completing the task. 

\subsection{Experiment Procedure }
Participants were randomly assigned to one of the two visualization conditions and a block order (e.g., public visualization first). In the first block of trials, participants performed a practice task as preparation for the trials where no information was present. They then did the no-information trial for the first block. Participants next did a practice task for the information (visualization) conditions in the first block. They then completed the nine visualization trials in the block. After completing the first block, participants were informed of how the trials in the second block differ from the first; i.e. they are either told that only they will have access to the data, or that others will also be able to view the visualizations to make their decisions. Then, they completed the trial where no prior information was provided followed by the visualization trials. Between the blocks, we included an attention check question that asked the participant whether they had just completed a block where all participants saw the visualization or not. This attention check was used to filter out participants who did not understand the information access condition.

To better understand how reasoning strategies changed with private vs.\ public visualization and the presence of a visualization, we asked participants to describe the strategy they used to select their preferred location. We asked participants to provide a strategy in a total of four trials: the two no-information trials (one in each block) and the final visualization trial in each block. On the final page of the study, we obtain demographic information (age, gender) and ask participants two additional attention check questions about the task (Selecting a location) and the number of simultaneous decision-makers in the task (30). These questions were combined with the first attention check question to filter participants who didn't show evidence of paying attention. 

\subsection{Participants}
We recruited U.S.-based participants on Amazon Mechanical Turk who had a HIT acceptance rate of 97\% and had completed at least 300 tasks. We aimed to recruit at least 375 participants, which powered the experiment to detect an effect size of 10\% with 80\% power for the within-subject effect of information access.

\subsection{Expectations}
In the private information setting, the participants pick a location under the assumption that other players will be drawn from the same distribution as the visualized distribution. Their best response is to select the visualized higher payoff location. There is also a ground truth response to the estimated probability that their preferred location receives a  higher payoff in the private visualization setting. This estimate is calculated using the binomial distribution for 30 observations and $p$ equal to the visualized proportion selecting A or B (depending on which the participant selected).

In the public visualization setting where participants must anticipate others' behavior, there is no ground truth response. We expect the proportion of participants selecting the location with a higher payoff in the visualization to decrease as some participants anticipate others' reactions. It is also natural to expect participants' estimates of the probability of having selected the higher payoff location to deviate more from the ground truth defined on the visualized data alone.


\begin{figure*}[t]
 \centering
  \includegraphics[width=\linewidth]{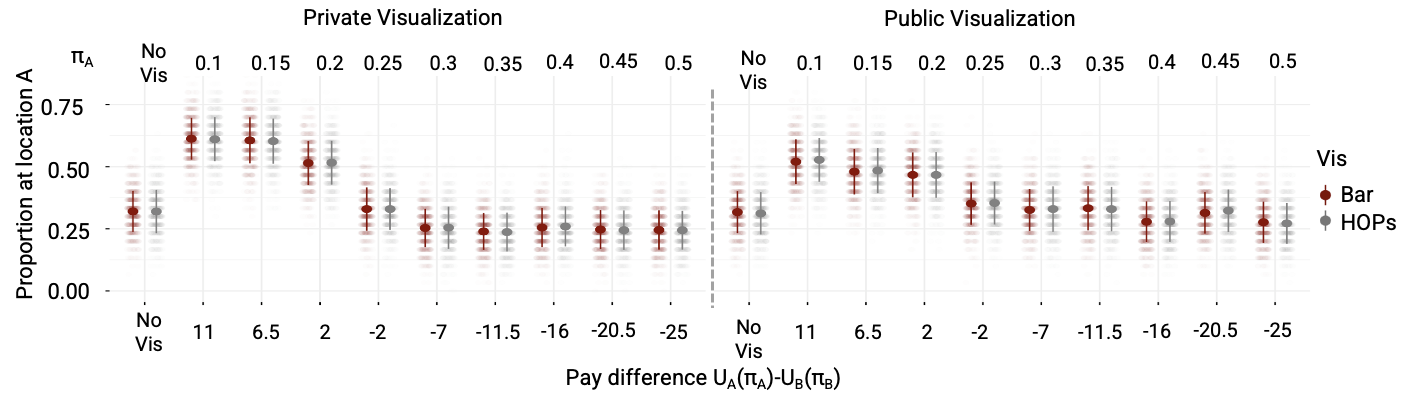}
   \caption{ Aggregate proportion of people selecting location A per visualization and information Condition in 1000 bootstrap samples groups of n=30 observations with bootstrap 95\% percentile confidence intervals.}
   \label{fig:proportionsummary}
 \end{figure*}


\section{Results}
\subsection{Data Preliminaries}
We recruited 606 participants in total. After excluding participants who failed to pass the attention check questions about the information access condition in the first block of trials they performed, as well as the two additional questions about the number of simultaneous decision-makers and the type of decision task, we were left with 400 participants. Regardless of exclusions, participants received a base payment of \$2 for 13 minutes of work on average. In addition to the base, payment participants received an average bonus of \$1.57. 

To calculate the payoffs in the private visualization condition, for each trial, we generated a sample of 29 responses from a binomial distribution with probability $p$ corresponding to the visualized proportion of participants at location A ($\prA$). 
To generate the payoffs in the public visualization condition, we randomly paired the participant with 29 other participants who saw the same visualization condition and computed their payoffs based on the proportion of people selecting location A and location B in this participant sample. To provide robust estimates of the probability of selecting location A, $\prA$, for our analysis, we resample with replacement 30 participants 1000 times for each combination of visualization type, information signal (proportion), and information access (private/public visualization).  

\subsection{Visualization Equilibrium}


\subsubsection{Overview of results}
\Cref{fig:proportionsummary} presents the results of the private (left) and public (right) visualization conditions, with point estimates showing bootstrapped proportion estimates and bootstrapped 95\% confidence intervals. Participants' decisions of what location to choose in both the private and the public visualization condition vary with information signal (proportion) changes, with more people selecting location A when the visualized proportion implies higher payoffs for selecting location A.

Recall that in the private visualization condition, the best response is to choose the higher payoff location according to the visualization, which is A for proportions less than $\prAnash \approx 0.22$. Surprisingly, a considerable proportion of participants (25\% to 40\% depending on visualized signal) do not best respond to the visualization. There are several possible explanations for this. Some participants may have failed to recognize that using the sample estimate of $\prA$ was their best bet under these conditions, instead treating the task like a gamble and choosing the lower payoff outcome according to the visualization because they thought they might get lucky. Or, participants might have misinterpreted the conversion table, which showed a distribution of payoffs of A and B under different proportions, to imply that all proportions were equally likely irrespective of the visualization. Our analysis of strategy descriptions below sheds some light on this.   

Recall that we included one trial in each block of trials where no prior information was present (\Cref{fig:proportionsummary}, far left of each subplot). As expected, we see no observable difference between these no-information trials between the public and private visualization conditions, as the no-information task is nearly identical. We observe realized proportions closest to those observed in the no-information trials at a visualized proportion of 0.25 for the private condition and 0.45 for the public visualization condition, though 0.3 and 0.35 are very close as well. Comparing the left to the right side of \Cref{fig:proportionsummary}, the results from the public visualization condition resemble a more compressed (toward 50\%) version of the private visualization conditions, meaning that with public visualizations, participants chose the higher payoff location according to the visualization less. We expect this result since public access to the visualization adds ambiguity in that the best decision is no longer directly implied by the visualization. Static versus hypothetical outcome bar charts appear to yield very similar observed proportions in both the private and public conditions suggesting that visualization effects on aggregate outcomes are small.    

Recall that we are interested in comparing strategic behavior resulting from visualizations with strategic behavior according to Nash equilibria, the standard prediction of game theory (\Cref{sec:nash}), and the socially optimal outcome (\Cref{sec:social}). When the information signal shows approximately equal payoffs at locations A and B (corresponding to the Nash Equilibrium of the game) agents split across the locations roughly equally. From \Cref{fig:proportionsummary} right, the observed proportion is closest to the Nash optimal proportion (0.22) when the visualized proportion is 0.4 and 0.5. We observe behavior closest to the socially optimal proportion (0.44) when the visualized proportion is 0.15 and 0.2. In the public information condition, visualizing proportions lower than Nash (0.22) obtains higher social welfare. 

\subsubsection{Estimating Visualization Equilibria}
Recall that at the visualization equilibrium, $\prAviseq = \STRAT(\VIS(\prAviseq))$, or in other words, the observed proportion $\prA$ choosing location A mimics the visualized proportion $\prApred$. We calculate a precise estimate of the visualization equilibrium for static bars versus HOPs in the public condition using a linear regression model that estimates the proportion of people who go to location A given the visualization condition and the visualized proportions: 

\begin{equation}
\begin{split}
    \text{Observed Proportion at A} \sim \text{Visualization Condition} \\
    +\text{Visualized Proportion A}
\end{split}
\end{equation}

The visualization fixed point (equilibrium) occurs when the visualized proportion is 0.34 [0.32,0.36] for Static Bar Chars and 0.35 [0.33,0.37] for HOPS. These highly overlapping estimates are not surprising given that our experiment was not necessarily powered to detect small differences between private versus public information conditions at an aggregate level. The equilibrium for both visualization conditions is located between the Nash ($\prAnash=0.22$) and the Social Optimum ($\prAopt=0.44$). Specifically, in the visualization equilibrium, participants show less extreme behavior, in aggregate, than in the Nash equilibrium, and higher social welfare.


\subsection{Modeling Individual Decisions \& Judgments}

As described above, the compression toward 50\% in Figure \ref{fig:proportionsummary} right relative to the left shows that relative to private visualization, participants became less likely to select the visualized higher payoff choice with a public visualization. This suggests visualization anticipation, or a change in behavior when others have access to a visualization of payoff relevant information.

\begin{figure*}[t]
 \centering
  \includegraphics[width=\linewidth]{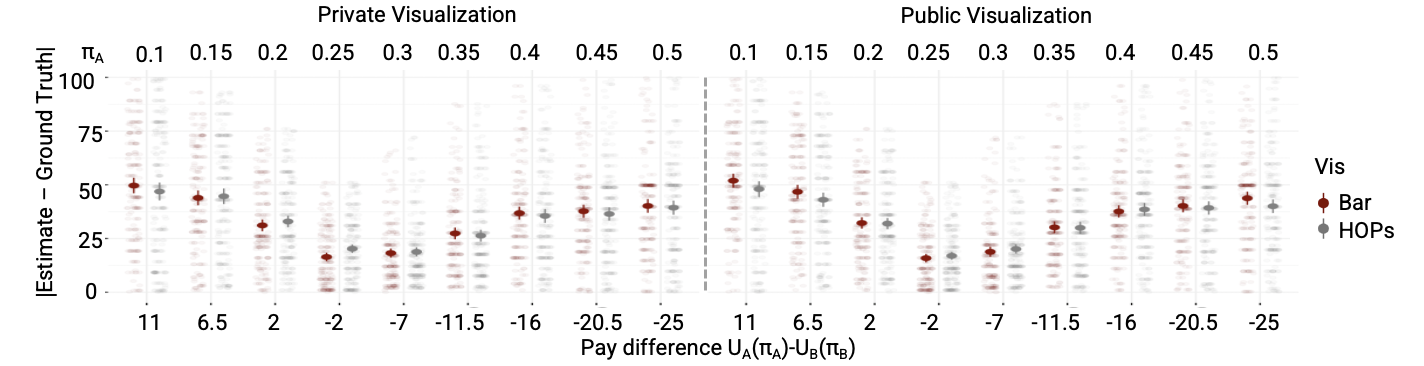}
   \caption{ Absolute difference between reported probabilities and ground truth probabilities  by information condition, visualization type and visualized signal (proportion).}
   \label{fig:difffromtruth}
 \end{figure*}

The probability responses participants provided on each trial provide more evidence of anticipation. Figure~\ref{fig:difffromtruth} reports absolute error in the probability estimates relative to the ground truth defined for the private visualization condition for all trials. Estimates varied significantly with the information signal and slightly depending on whether the visualization was public or private. Overall, we see high amounts of error for some proportions, specifically those where the difference in payoffs is higher, and especially when A is the correct location. Errors are minimized, though still close to nearly 20 percentage points on average, when the visualized proportion is 0.25. 
In the trials where the error is greatest, the correct response was close to 0 or 100\%, suggesting that participants perceived the probability as closer to 50\% than they should have for most trials. 

While the probability estimates were similar across public and private visualization conditions, we noted slightly more shrinkage toward 50\% in the reported probabilities in the public visualization condition, suggesting that participants were sensitive to the additional uncertainty around what other participants would do with the visualization. Again we see few observable differences between bar charts and HOPs.  

Below we specify regressions on binary decisions and probability estimates to better understand the differences between public versus private visualizations, information signals, and visualization conditions. 

\subsubsection{Modeling Binary Decisions}

We specify a mixed-effects logistic regression to predict whether a participant will choose the visualized higher payoff (i.e., Best Responded below, dummy coded as 1 if they chose the location and 0 if not). We include a dummy variable for visualization condition (Visualization), for whether the visualization was public versus private (Information Access), for whether location A or B was visualized to have high payoff (Visualized High Payoff Location), and for block order (Block Order). We include the absolute difference in expected payoff between location A and B (Abs. Visualized Payoff Difference =$\mid\payoffA(\prA)-\payoffB(\prB)\mid$) according to the visualized proportion as a continuous variable. We include an interaction term (\text{Visualization} * \text{Information Condition}) to measure if the rate of best response varies as a joint function of the visualization and the information condition. Finally, we include random intercepts per participant to account for individual differences among participants. The model specification was chosen a priori to include only the interaction in which we were interested (between visualization approach and information access). 

\begin{equation}
\begin{split}
   \text{Best Responded} &\sim \text{Visualization} * \text{Information Access} \\
   &+ \text{Visualized High Payoff Location } \\
   &+ \text{Abs. Visualized Payoff Difference} \\
   &+\text{Block Order}+1\mid\text{Parcipant} 
 \end{split}
 \end{equation}

\begin{figure}[t]
  \centering
  \includegraphics[width=\linewidth]{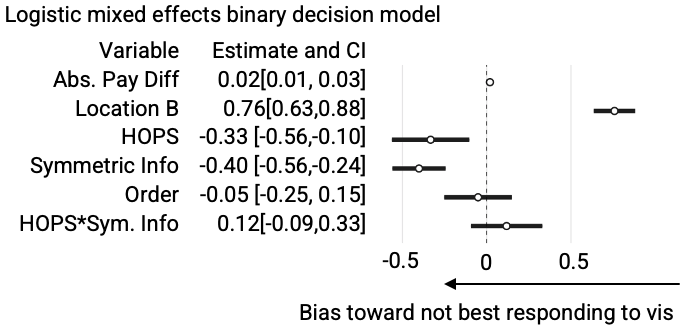}
  \caption{Coefficients (in log odds probability of choosing the visualized higher payoff location) with 95\% confidence intervals from a regression on participants' binary decisions.}
  \label{fig:binaryanticipation} 
\end{figure}%

Results are shown in  Figure \ref{fig:binaryanticipation}. We use the 95\% CIs on the coefficients (and whether they include 0) as a rough metric for whether an effect is well supported or not. The model intercept for the rate of best responding to the visualization is 0.41 in log odds space, corresponding to a probability of roughly 60\% percent chance of selecting the visualized higher payoff location. As we expected for visualization anticipation, when the visualized information is public, participants are 0.67 times (-0.41 log odds) as likely to choose the visualized higher payoff location. Participants who saw HOPs were 0.71 (-0.33 log odds) times as likely to select the visualized higher payoff location relative to participants who saw bar charts. 
While we can only speculate, one possible reason might be that bar charts present point estimates without a visual depiction of uncertainty, making it easier for participants to feel confident about their choice without the complexity of an uncertainty encoding. 

To understand this effect in light of public versus private information, we must consider the interaction term, which indicates a slightly bigger difference in probability of choosing the visualized higher payoff location in going from private to public visualization with static bar charts relative to HOPs (Figure \ref{fig:interactioneffects}). However, this interaction effect is unreliable (0.12; 95\% CI: [-0.09, 0.33]). While the point estimate is in line with our expectation that HOPs, as a more salient uncertainty depiction, make it more difficult to anticipate how others will respond, that the estimate is unreliable speaks to the relatively random decision making we observed across our experiment.

Every unit change in the absolute pay difference, $\mid\payoffA(\prA)-\payoffB(\prB)\mid$, is associated with a very small (2\%; 0.02 log odds) but reliable increase in the odds of selecting the visualized higher payoff location. Participants were 2.1 times (0.76 log odds) as likely to select the visualized higher payoff location when it was B. This relatively large effect is unexpected; however, we suspect that some participants reacted to the expected value of the payoffs irrespective of the visualization, which would imply that location B is the better location. Our analysis of reported strategies (Section 5.4) below corroborates this.

\subsubsection{Modeling Probability Estimates}
We fit a similar mixed-effects model to participants' probability estimates. We first compute the absolute error in estimated probabilities by taking the absolute value of the difference between the empirical and the ground truth probability for the visualized higher payoff location (after first subtracting from 1 the probabilities of participants who did not choose the visualized higher payoff location).

\begin{equation}
\begin{split}
   \text{Abs. error in probability estimate} \sim \text{Visualization}*\text{Information Access}\\+ \text{Visualized High Payoff Location} + \text{Abs. Visualized Payoff Difference}\\
   +\text{Block Order}+1\mid\text{Participant} 
 \end{split}
\end{equation}

\begin{figure}[t]
  \centering
  \includegraphics[width=\linewidth,height=4.25cm]{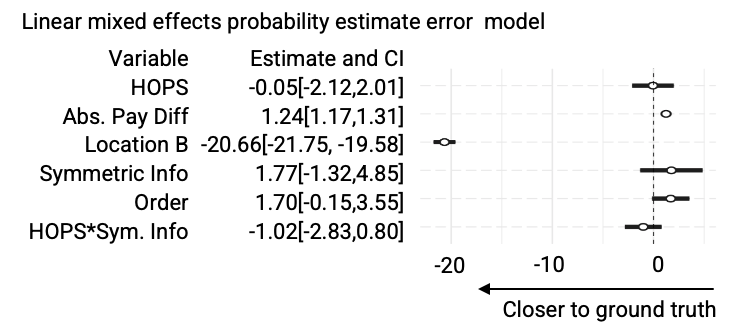}
  \caption{Coefficients with 95\% confidence intervals from a regression on participants' probability estimates.}
  \label{fig:uncertaintyanticipation}
\end{figure}

Results are shown in Figure ~\ref{fig:uncertaintyanticipation}. The model intercept for the absolute error is 32.4 percentage points. For each unit change in the visualized absolute payoff difference, $\mid\payoffA(\prA)-\payoffB(\prB)\mid$, the absolute difference between the empirical and the ground truth probability increases by 1.2 percentage points. The probability estimates indicate slightly greater error (about 2 percentage points on average) for changing from private to public visualization, but this effect is not reliable. This result may suggest that the probability estimate question was not well necessarily understood by participants, or that participants perceived greater uncertainty across the board than they should have. As noted above, we observe a general compression effect in probabilities across the board, toward 50\%, and for several trials where location B is the higher payoff location according to the visualization, the ground truth probability defined for the private condition is approximately 50\% $(\prA=0.2)$ and 72\% $(\prA=0.3)$, so that participants appear to be much more accurate. When the visualized higher payoff was location B, the absolute difference between the empirical and the ground truth probability decreases by a much larger amount: 20 points, which may be an artifact of many participants responding based on the payoffs across trials rather than the visualized information.

\begin{figure*}[t]
  \centering
  \includegraphics[width=\linewidth]{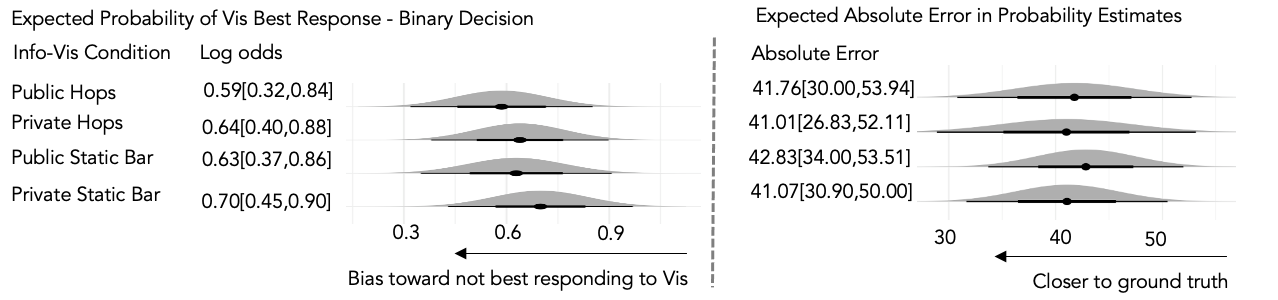}
  \caption{Model-estimated probability of best responding to the visualization (binary decision; left) and absolute error in probability estimates (probability judgment; right), marginalizing over other experimental manipulations.}
  \label{fig:interactioneffects} 
  
\end{figure*}

The results show no clear difference in probability estimates using HOPs relative to using static bar charts (-0.05; 95\% CI [2.1, 2.0]. This effect is somewhat surprising, as participants can use HOPs directly to estimate the ground truth probability in private trials, provided they remain aware of the payoff transformation on the visualized outcome (which the text above the chart provided). Again to fully understand the effects, we must consider the interaction term. Again we see a slightly bigger difference in conditional means for static bars when going from private to public than we see from HOPs, but in light of the overall magnitude of error this effect is small and unreliable (Figure~\ref{fig:interactioneffects}). 
The poor results for HOPs in the private trials appear consistent with a recent prior crowdsourced study with a similarly complex experimental task that found that most participants failed to ascertain that HOPs conveyed probability of superiority directly if one used them to estimate the number of draws where one group had a higher value~\cite{KaleAlex2020VRSf}. 
 
\subsection{Strategies}
Recall that we asked participants to describe the strategies they used in four trials: the two no-visualization trials in each block and the final visualization trial in each block. We qualitatively analyzed the reported strategies to understand how participants made decisions, focusing on the public visualization condition where anticipation plays a role.

\subsubsection{Reported Strategies}
Overall we collected 1600 strategy responses. We filtered out responses that were not informative, such as ``n/a" or ``I looked at the table". We performed analysis on the 1,163 remaining responses. The first author developed a coding scheme using a top-down approach based on iterative passes through the strategies, then the second author reviewed codes, with ambiguities resolved through discussion. The strategy codes are not mutually exclusive and some participants reported using multiple strategies. 

We report full results in supplemental material, focusing here on participants in the public visualization case, where the best response strategy was not well defined.
Our analysis coded for the following:



\begin{table}
\caption{Frequency of strategies in the public information setting}
\centering
\begin{tabular}{>{\quad}lcccc}
\toprule
\rowgroup{Information Access} & \multicolumn{3}{c}{Public}\\
\midrule
\rowgroup{Strategy} &No Vis& Bar& HOPS\\
\midrule
\rowgroup{Anticipation}\\
Visualization Anticipation & NA & 32\% & 24\%\\
Payoff Anticipation & 28\% & 1\% & 2\%\\
\midrule
 \rowgroup{Random} & 11\%&9\%& 6\%\\
 \rowgroup{Reacting Directly} &NA & 42\%& 53\% \\
 \rowgroup{Reacting to Payoffs} &61\% & 15\% & 15\%\\
\bottomrule
\end{tabular}
\vspace{-1em}
\end{table}

\noindent \textbf{Random selection}:  9\% of participants using Bar Charts and 6\% of participants using HOPs reported selecting randomly between the two locations. Some implied they could not make inferences about the distribution of other players' decisions and thus believed the outcomes would be a \textit{``coin toss."} Participants reported using randomization as a strategy slightly more when no information was visualized (11\% relative to 9\% with Bar Chart and 6\% with HOPs).

\noindent \textbf{Responding to the Payoffs}:
61\% of participants in the no visualization condition reported making decisions by responding to the payoff distribution. Participants noted, for example, that location B had \textit{``higher floor (22 vs 11)"}  and \textit{``higher ceiling (80 vs 40)"} relative to location A, and thus selected location B. Another participant noted that location A had \textit{``lower points reduction,"} and thus they selected location A as if to minimize loss. A smaller proportion (15\%) reported responding to payoffs in both visualization conditions. 

\noindent \textbf{Anticipation}: 32\% of participants using static bar charts and 24\% of participants using HOPs reported using visualization anticipation as a strategy; that is, anticipating how other participants may make decisions in light of the visualization and adjusting their behavior to account for this. For example, participants said \textit{``As Location A seems to win more in this scenario, I predict more Mturkers will pick location A, making B ultimately more profitable."} The slightly larger proportion mentioning anticipation for bar charts than HOPs might reflect how bar charts, by not directly encoding uncertainty, are less ambiguous and therefore easier to imagine reactions to. In the no visualization trial, 28\% of participants reported reasoning about how other players may react to the payoffs (Payoff Anticipation). A much smaller proportion of participants who saw HOPs and Bar charts reported reasoning about how other participants would respond to the payoffs irrespective of the visualization e.g., \textit{``Even though everyone has access to the same visual data, I don't think I can assume that everyone would automatically choose Location B, and the number of people I need to pick Location A is not too high a percentage to make the risk worthwhile."}

\noindent \textbf{Best Responding to Visualization}: 42\% of participants using bar charts reported best responding to the visualization by selecting the location which was visualized to have a higher payoff, and 53\% reported best responding to the visualization when the visualization was HOPs. One participant stated \textit{``Again, my strategy was just based on the visualization. In most, but not all cases, the B choice showed a higher payoff, so that's the one I picked"}. This result might corroborate the anticipation code results since anticipation and best responding imply opposite strategies. 


\section{Discussion}
Not surprisingly, we find evidence that different information signals and visualizations result in agents making decisions at different rates. 
Our results provide insight into how untrained participants respond in a congestion game scenario, and how anticipating others' reactions to a visualization appears to change behavior in a strategic setting.


\textbf{Visualized signal matters.} Visualizing different signals resulted in participants making decisions at different rates across trials. While this is not surprising, we found that with public visualizations, visualizing signals close to a Nash equilibrium for the game does not result in the realization of the equilibrium. Instead, participants were more indifferent to the decision of location than they should have been, with approximately 50\% going to each.

\textbf{The visualization equilibrium lies between the Nash equilibrium and socially optimal value}. Across visualization conditions, participants obtain more of the socially optimal welfare relative to Nash. However, an important consideration is that for our game, the social optimum was also closer to 50\% than the Nash equilibrium. Future work should vary the game to gain further intuition into the potential relationships between the visualization equilibrium, Nash equilibrium, and socially optimal value. 


\textbf{Anticipation reduced "following the visualization."}
Participants were only about two-thirds as likely to best respond to the visualization when others had access to it.
Multiple participants reported reasoning about player distributions and making statements about what proportion of other players they thought would choose a location, e.g., ``I was going for, B has many more options for a higher payoff. The problem of course is that everyone else knows this too, so if we all choose B, it's not a higher payoff.  Chose B anyway.''
A future direction well-motivated by these results is to apply behavioral models like cognitive hierarchy~\cite{stahl1993evolution} or level k~\cite{camerer2004cognitive}, both of which assume that players vary in their sophistication when it comes to anticipating others' decisions. For example, in a level k framework~\cite{camerer2004cognitive}, a level one player best responds to random play, whereas a level two player best responds to a mix of random play and level one play. Such models are used to estimate the distribution of levels in a population, which we suspect may be influenced by the visualization interface.

\textbf{Participants' perceived their decisions as more random than they were.} Probability estimates were compressed toward 50\% across trials and varied slightly, but not reliably, between the private and public information condition. The latter was somewhat surprising as we expected these estimates to reflect their greater uncertainty about the aggregate outcome when the visualized information was public. There are various reasons this might be the case, which we discuss further below in Limitations (Section 6.1). 

\textbf{HOPs reduced "following the visualization."} Participants were less likely (0.71 times) to choose the visualized higher payoff location with HOPs overall.
 We had hypothesized that a visualization that makes uncertainty affecting the signal more salient would make it harder for people to anticipate others' reactions. If most people respond to others viewing the visualization by changing their response from the private condition to instead choose the lower payoff location according to the visualization, as some participant strategies suggested, than we would expect a bigger difference in decisions for static bars between private and public than for HOPs.  
However, while we do see slightly larger differences in both probability of selecting the visualized higher payoff location and in probability estimates between private and public static bars relative to HOPs, corroborated by our strategy analysis, these effects are small and unreliable, perhaps as a result of the tendency for compression toward 50\% in both binary decisions and probability estimates. 

\subsection{Limitations}
Our study considered the trivial information-theoretic setting in which study participants were only uncertain about the behavior of other agents and not the payoff relevant state. The canonical information design question considers settings in which there is uncertainty about the payoff relevant state. In such settings, the agents do not know the payoff structure a priori and instead learn about the unknown state through Bayesian updating.  Agents are assumed to update like Bayesians when presented with information; however, this assumption is inconsistent with the research in Judgment and Decision Making ~\cite{tversky1974judgment} that suggests that people often rely on simple heuristics when making a decision. Visualizations have been shown to improve Bayesian reasoning and mitigate cognitive bias by reducing over-reliance on heuristics. In a setting where payoff structures are not known a priori, visualizations can act as a communication mechanism to support statically coherent updating.

Knowledge about the payoffs, emphasized through interface design choices such as the conversion table and text formatting, may have biased participants' responses leading them to react to the expected value of the payoffs irrespective of the visualization.  Furthermore,  participants were required to translate the visualized information through a payoff function resulting in an additional cognitive load that may have led participants to rely on the expected payoff as a decision-making heuristic.

Our choice to incentivize only the binary decision problem and not the probability estimate may have led participants to decouple the probability question from the overall task. In contrast to the binary decision responses, the probability estimate responses are robust to the information condition. It seems possible that participants responded more to how confident they felt about the tasks in general or based on individual risk appetites. Prospect theory \cite{tversky1992advances} provides a possible explanation for the observed estimates by suggesting that participants may not directly report the probability of an event and instead report the weighted probability, reflecting risk attitudes resulting in decision weights that are generally lower than corresponding probabilities. Linear-in-log-odds models could be useful in the future to provide additional insight into bias in probability estimation~\cite{zhang2012ubiquitous,gonzalez1999shape}.

\subsection{Applying the Vis Equilibrium Solution Concept}
Our work conceptualized a new solution concept, visualization
equilibrium, to study strategic environments in which payoff relevant
information is communicated to agents using a visualization. The
visualization equilibrium is a fixed point where the visualized
behavior equals the realized behavior.  Visualizations that are not at
the visualization equilibrium may result in unpredictable dynamics in
strategic environments.  Visualization equilibrium is an empirical
concept and to employ visualization equilibrium, e.g., in navigation
applications, the behavioral responses to the visualizations are
learned from data so that the visualization equilibrium can be
identified and obtained.
\section{Conclusion}
We introduced the concept of visualization anticipation, visualization equilibrium and explored the effects of different ways of visualizing payoff-related information to agents in a strategic decision-making task. We found that anticipation accounts for a significant change in behavior among individuals in strategic settings. These results apply to numerous situations where payoff relevant information is visualized to all agents, from online ad bidding marketplaces to transit apps to presenting results of political forecasts. The visualization equilibrium, was similar for the visualization types we explored. At the visualization equilibrium agents displayed less selfish behavior thus they obtained more social welfare.



\bibliography{biblio}
\end{document}